\documentclass[twocolumn,prl,showpacs]{revtex4}

\usepackage{graphicx}
\usepackage{dcolumn}
\usepackage{bm}
\usepackage{amsmath}
\usepackage{amssymb}

\begin{document}

\title{Long-range ferromagnetism of Mn$_{12}$ acetate single-molecule magnets
under a transverse magnetic field}

\author{F. Luis$^{*,1}$, J. Campo$^{1}$, J. G\'omez$^{2}$, G. J.
McIntyre$^{3}$, J. Luz\'on$^{1}$, and D. Ruiz-Molina$^{2}$}
\address{$^{1}$Instituto de Ciencia de Materiales de Arag\'on,
CSIC-Universidad de Zaragoza, 50009 Zaragoza, Spain.\\
$^{2}$Institut de Ciencia de Materials de Barcelona, Campus de la
UAB, Bellaterra 08193, Spain.\\
$^{3}$Institut Laue Langevin, 6 rue Jules Horowitz, 38042
Grenoble, France.}


\begin{abstract}
We use neutron diffraction to probe the magnetization components
of a crystal of Mn$_{12}$ single-molecule magnets. Each of these
molecules behaves, at low temperatures, as a nanomagnet with spin
$S=10$ and strong anisotropy along the crystallographic $c$ axis.
Application of a magnetic field $B_{\perp}$ perpendicular to $c$
induces quantum tunneling between opposite spin orientations,
enabling the spins to attain thermal equilibrium. Below $\sim 0.9$
K, intermolecular interactions turn this equilibrium state into a
ferromagnetically ordered phase. However, long range ferromagnetic
correlations nearly disappear for $B_{\perp} \gtrsim 5.5$ T,
possibly suggesting the existence of a quantum critical point.
\end{abstract}

\maketitle


Magnetic nanostructured materials have opened new frontiers for
basic science and applications, due to their unique size-dependent
properties and the emergence of quantum phenomena
\cite{Chudnovsky98}. Several fundamental problems remain, however,
to be understood. Of particular interest are the observation of
phase transitions induced by mutual interactions between
nanoscopic magnets in dense arrays. A crucial question here is the
mechanism by which the ordered equilibrium phase is attained. For
very small magnetic clusters, zero-point quantum fluctuations
(e.g. quantum tunneling) are expected to dominate the relaxation
process at very low temperatures \cite{Evangelisti04}. Eventually,
these fluctuations might even suppress long-range order provided
they become sufficiently strong against both interparticle
interactions and decoherence \cite{Sondhi97}. Such a quantum
critical point has been observed for 3-D model Ising ferromagnets
\cite{Bitko96} but its existence for arrays of nanomagnets remains
uncertain. Understanding the interplay between ordering and
quantum fluctuations can be important for applications of these
nanomagnets in ultrahigh density magnetic recording and quantum
computation, when it is necessary to control the quantum behavior
of entangled interacting qbits.

Usually, however, these phenomena are masked by the particle's
size distribution and disorder present in even the most
homogeneous samples \cite{Sun00}. By contrast, molecular magnetic
clusters \cite{Christou00,Sessoli03} are ideal candidates for
these studies \cite{Fernandez00,M-Hidalgo01}. In the case of
Mn$_{12}$ acetate \cite{Lis80}, the first and most extensively
studied member of the family of single-molecule magnets, clusters
contain $12$ manganese atoms linked via oxygen atoms, with a
sharply-defined and {\em monodisperse} size. At low temperatures,
each of them exhibits the typical behavior of a magnetic
nanoparticle, such as slow magnetic relaxation and magnetization
hysteresis loops, due to the combination of an $S=10$ magnetic
ground state with appreciable magnetic anisotropy. And finally,
they organize to form tetragonal molecular {\em crystals}. Since
molecular spins couple via dipolar interactions, these crystals
are nearly perfect realizations, with magnetic units of mesoscopic
size, of the Ising quantum model.

Long-range magnetic order has however not been observed for
Mn$_{12}$ yet. The reason is that the spin reversal via resonant
quantum tunneling \cite{Friedman96,Hernandez96,Thomas96,Thomas99}
becomes extremely slow at low temperatures (of order two months at
$T = 2$ K). For the time scales $\tau_{\rm e} \sim 10^{2} -
10^{4}$ s of a typical experiment, the spins are unable to attain
thermal equilibrium below a blocking temperature $T_{\rm B} \sim
3$ K, which is higher than the ordering temperature $T_{\rm c}$.
It has also been argued \cite{M-Hidalgo01} that hyperfine bias
caused by randomly frozen Mn nuclear spins might hinder the
occurrence of long-range order in Mn$_{12}$. Here we circumvent
these experimental problems by the application of a transverse
magnetic field $B_{\perp}$ that promotes quantum tunneling of the
molecular spins. We report neutron diffraction data that point to
the existence of long-range ferromagnetic order in Mn$_{12}$ that
can be suppressed by either increasing temperature up to $T_{\rm
c} \simeq 0.9(1)$ K or magnetic field, above $\sim 5.5 (5)$ T.

Magnetic diffraction of thermal neutrons is a suitable tool for
these studies because neutrons can probe the magnetization along
the anisotropy axis \cite{Robinson00}, i.e. the order parameter,
and provide a very accurate determination of the crystal's
orientation with respect to the applied magnetic field. The $\sim
0.5 \times 0.5 \times 1.5$ mm$^{3}$ single crystal of deuterated
Mn$_{12}$ acetate, [Mn$_{12}$ (CD$_{3}$COO)$_{16}$ (D$_{2}$O)$_{4}
$O$_{12}$] $\cdot 2$ CD$_{3}$ COOD$\cdot4$D$_{2}$O, was prepared
following the original method of Lis \cite{Lis80}. It was glued to
a copper rod in good thermal contact with the mixing chamber of a
$^{3}$He-$^{4}$He dilution refrigerator, which gives access to the
temperature range $45$ mK $\leq T \leq 4$ K. The {\em c} axis was
carefully oriented to be perpendicular to the vertical field $0
\leq B_{\perp} \leq 6$ T applied by a superconducting magnet. From
the orientation matrix measured at zero field and $T=4$ K, we
estimate that the crystallographic $(\bar{1} 1 0)$ direction lay
within $0.1(1)$ degrees of the magnet axis.

\begin{figure}[b]
\includegraphics[width=8.5 cm]{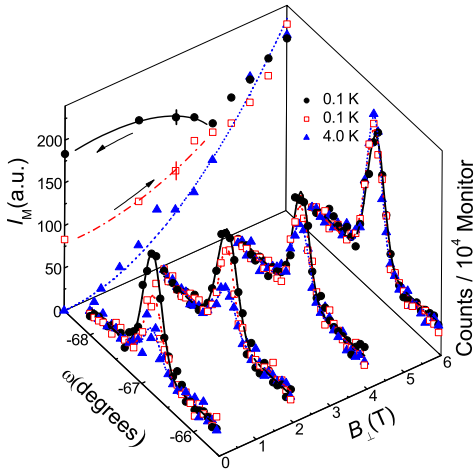}
\caption{\label{ImvsB} (Color online) Rocking curves for the
$(\bar{2} \bar{2} 0)$ reflection measured at two different
temperatures and four magnetic fields. The counting statistics are
typical for such a small crystal under these conditions. The lines
are Gaussian fits. Numerical integration of these rocking curves
gives the diffracted intensity $I$. The magnetic diffraction
intensities $I_{m}$ were obtained at each temperature and field by
subtracting from the total intensity the value measured at $4$ K
and $B_{\perp} = 0$. $\vartriangle$, $T=4$ K; $\bullet$ and
$\circ$, $T=100$ mK measured while increasing and then decreasing
$B_{\perp}$, respectively. The dotted line is a least squares fit
to a parabola $A B_{\perp}^{2}$.}
\end{figure}

At any temperature $T \leq 4$ K, we measured a series of Bragg
diffraction reflections as a function of $B_{\perp}$
\cite{neutron}. Each reflection $(h k l)$ contains a nuclear
$I_{\rm N}$ contribution and a magnetic $I_{\rm m}$ contribution.
The former contains information about atomic order, whereas the
latter is proportional to the square of the magnetization
components perpendicular to the $(h k l)$ direction. The nuclear
contribution can be obtained by measuring the intensity at zero
field in the paramagnetic phase (see Fig. \ref{ImvsB}). By
subtracting this, $I_{\rm m}$ can be estimated at any field and
temperature.

Given the strong anisotropy of Mn$_{12}$, the magnetization must
be confined in the plane defined by the anisotropy axis $c$ and
the magnetic field, with components $M_{z}$ and $M_{\perp}$
respectively. The experimental protocol followed for each
reflection line is illustrated in Fig. \ref{ImvsB}, where we plot
raw rocking curves obtained for the $(\bar{2} \bar{2} 0)$
reflection. For this example, the momentum transfer is orthogonal
to both the magnetic field and the anisotropy axis. In addition,
it has a very small nuclear contribution. Therefore the diffracted
intensity must be sensitive to both $M_{z}$ and $M_{\perp}$
components. At $4$ K, $I_{\rm m} \propto B_{\perp}^{2}$, as
expected, since $M_{z}=0$ in the paramagnetic state and
$M_{\perp}$ is proportional to $B_{\perp}$. At $100$ mK, by
contrast, a large additional contribution to $I_{\rm m}$ shows up
in the low-field region (for $B_{\perp} < 5$ T). Since $M_{\perp}$
is nearly independent of $T$ (see the inset of Fig. 2), the
additional magnetic diffracted intensity reflects the onset of a
non-zero $M_{z}$. Furthermore, this low-$T$ contribution shows
hysteresis. Indeed, as shown in Fig.\ref{ImvsB}, $I_{\rm m}$ data
measured while {\it increasing} $B_{\perp}$ after the sample was
cooled at zero field from $T \simeq 1$ K, lie clearly below those
measured while {\em decreasing} it, merging approximately at
$B_{\perp} = 4(1)$ T. The hysteresis means that spins can attain
equilibrium within the experimental time $\tau_{\rm e} \simeq 7
\times 10^{3}$ s only above $B_{\perp} = 4$ T. The fact that this
field-induced "jump to equilibrium" happens at approximately the
same field for $T=100$ mK and $T=600$ mK confirms that relaxation
to equilibrium proceeds via temperature-independent tunneling
processes \cite{Luis00,Chiorescu00}.

\begin{figure}[t]
\includegraphics[width=7.6cm]{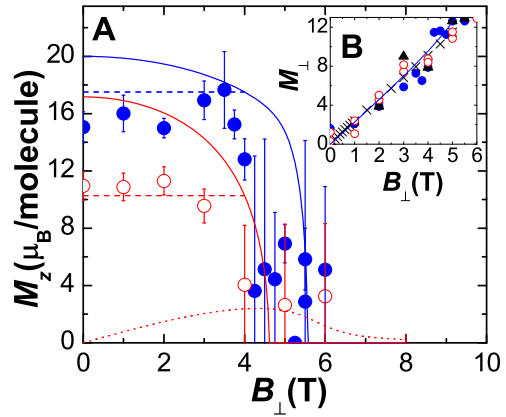}
\caption{\label{MvsB} (Color online) {\bf A} Longitudinal
magnetization $M_{z}$ of Mn$_{12}$ acetate measured while
decreasing the transverse magnetic field from $6$ T. $\bullet$ and
$\circ$ are for $T=47$ mK and $600$ mK, respectively. Solid lines
are calculated using Eq. (\ref{Hamiltonian}) and the parameters
given in the text. Horizontal dashed lines show $M_{z}$ calculated
at $4$ T, below which spins are 'frozen' by the anisotropy energy
barriers. The dotted line shows $M_{z}$ induced at $T=600$ mK by
the small misalignment of the crystal in the case of no
interactions ($J_{\rm eff}=0$ in Eq. (\ref{Hamiltonian})). {\bf B}
$\times$ perpendicular magnetization $M_{\perp}$ obtained at $T=4$
K with a SQUID magnetometer. Data obtained from neutron
diffraction are also shown: $\bullet$, $T=47$ mK; $\circ$, $T =
600$ mK; $\blacktriangle$, $T=4$ K.}
\end{figure}

\begin{figure}[t]
\includegraphics[width=7.6cm]{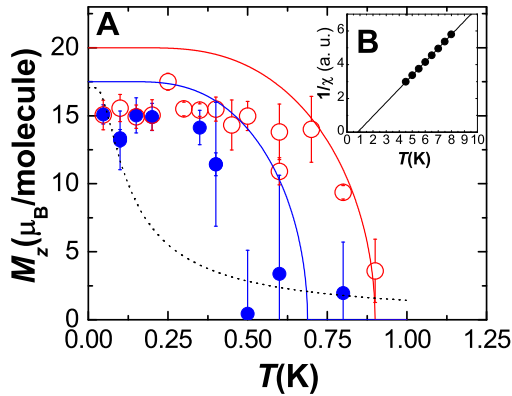}
\caption{\label{MvsT} (Color online) ({\bf A}) Longitudinal
magnetization $M_{z}$ obtained from neutron diffraction data
measured at $B_{\perp} = 0$ ($\circ$) and $4$ T ($\bullet$).
Measurements were recorded after cooling the sample in zero field
and subsequently applying $B_{\perp} = 6$ T before setting the
final field. The dotted line shows $M_{z}$ arising from the
maximum misalignment ($0.1$ deg.) of the crystal at $B_{\perp}=4$
T. Solid lines are calculations (for perfect orientation) that
include interactions via the mean-field Hamiltonian
(\ref{Hamiltonian}). ({\bf B}) Reciprocal equilibrium
susceptibility measured at $T>4.5$ K along the $c$ axis. The line
is a least-squares linear fit, giving $T_{\rm c} = 0.8(1)$ K.}
\end{figure}

To obtain $M_{z}$ and $M_{\perp}$ as a function of magnetic field
(Fig. \ref{MvsB}) and temperature (Fig. \ref{MvsT}), several
reflection lines were simultaneously fitted and the results
calibrated against SQUID magnetization measurements performed at
$4$ K. At our minimum temperature $T=47$ mK, $M_{z}$ is
approximately zero for $B_{\perp} \gtrsim 5.5(5)$ T and then it
increases when decreasing $B_{\perp}$, reaching $16 \mu_{\rm B}$
per molecule at zero field. The temperature dependence of this
zero-field $M_{z}$ is shown in Fig. \ref{MvsT}. It is
approximately constant until it begins decreasing sharply for $T
\gtrsim 0.6$ K. Above $1$ K, as at $4$ K, the fit gives $M_{z}
\sim 0$.

These experiments show without ambiguity the existence of a net
magnetization along the anisotropy axis and that it vanishes at
sufficiently high temperatures and perpendicular magnetic fields.
The qualitative resemblance between $M_{z}$ vs $B_{\perp}$ and
$M_{z}$ vs $T$ curves is typical of quantum ferromagnetic systems
\cite{Bitko96}. However, for such a strong anisotropy, any
deviation of the crystal from the perpendicular orientation can
induce, at $B_{\perp} > 4$ T, some magnetic polarization along the
anisotropy axis \cite{Chiorescu00} that would then remain frozen
below the irreversibility field ($4$ T). In our case, this
deviation is known from "high"-temperature diffraction data to be
smaller than $0.1$ degrees, which agrees fully with the small
$M_{z}$ values measured at $B_{\perp}=4$ T and above $500$ mK. In
Figs. \ref{MvsB} and \ref{MvsT} we show that such a small
misalignment cannot account, by itself, for $M_{z}$ observed at $T
> 100$ mK. In addition, for $T > 400$ mK, $M_{z}$ increases
significantly as the field is reduced to zero from $B_{\perp} = 4$
T, showing that, even at zero field, the spins tend to polarize as
they approach equilibrium. The existence of long-range magnetic
order is also supported by the finite temperature intercept
($0.8(1)$ K) of the reciprocal paramagnetic susceptibility shown
in the inset of Fig. \ref{MvsT}. We therefore conclude that
Mn$_{12}$ acetate is a ferromagnet for $T < T_{\rm c} \sim 0.9
(1)$ K and $B_{\perp} < B_{c} \sim 5.5(5)$ T. The ferromagnetic
nature of the ordered phase agrees with theoretical predictions
\cite{Fernandez00}, which however predict $T_{\rm c} \sim 0.45$ K,
lower than observed. In our view, the discrepancy might arise from
the fact that Monte Carlo calculations were performed for
point-like spins, whereas Mn$_{12}$ molecules are extended
nanoscopic objects \cite{dipolar}. It is worth mentioning that the
relatively strong hyperfine interactions do not prevent the
ordering of Mn$_{12}$ molecular spins probably because, as has
been observed recently \cite{Luis00,Evangelisti04}, Mn nuclear
spins also fluctuate rapidly when tunneling rates are sufficiently
fast.

These qualitative interpretation can be put on a solid basis with
the help of theoretical calculations. A simple way to introduce
interactions in the analysis is by making use of a mean-field
approximation
\begin{equation}
{\cal H} = -DS_{z}^{2} + C (S_{+}^{4}+S_{-}^{4}) - g\mu_{\rm B}
\left( B_{x}S_{x} + B_{y}S_{y} \right) - J_{\rm eff} \langle S_{z}
\rangle S_{z} \label{Hamiltonian}
\end{equation}
where $D$ and $C$ are the uniaxial and in-plane anisotropy
constants, $g=2$ the gyromagnetic ratio, $B_{x}$ and $B_{y}$ are
the magnetic field components along $a$ and $b$, $J_{\rm eff}$ is
a mean-field interaction parameter, and $\langle S_{z} \rangle =
M_{z}/g \mu_{\rm B}$ is the thermal statistic average of $S_{z}$.
We estimated $D = 0.62 k_{\rm B}$ by fitting the perpendicular
magnetization measured at $4$ K, while $C$ has been set to $2.5
\times 10^{-4}$ K in order to fit the critical field $B_{\rm c} =
5.5$ T. These are of the same order as the values obtained by
spectroscopic techniques \cite{Barra97}. We also set $J_{\rm eff}
\simeq 4.5 \times 10^{-3} k_{\rm B}$ to account for the
experimental $T_{\rm c} \simeq 0.9$ K.

\begin{figure}[t]
\includegraphics[width=7.6cm]{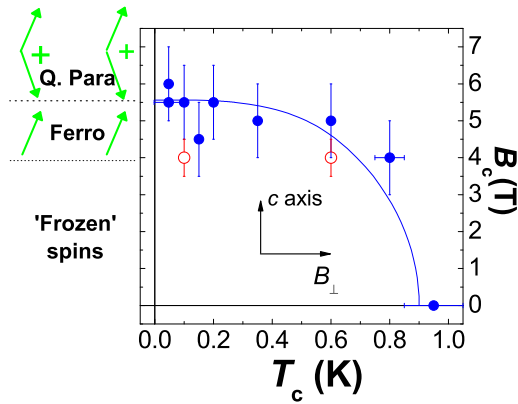}
\caption{\label{phasediagram} (Color online) Magnetic phase
diagram of Mn$_{12}$ acetate. The solid dots show the critical
magnetic field $B_{\rm c}$ at which the longitudinal magnetization
is observed to vanish at each temperature. The open dots give the
irreversibility field, below which spins do not attain equilibrium
within the experimental time. The solid line was obtained from
magnetization curves calculated in a mean-field approximation (Eq.
(\ref{Hamiltonian})).}
\end{figure}

We have calculated $M_{z}$ and $M_{\perp}$ by performing a
numerical diagonalization of Eq.(\ref{Hamiltonian}) followed by a
self-consistent calculation of the statistically averaged spin
components. Field- and temperature-dependent calculations account
reasonably well for $M_{z}$ and $M_{\perp}$, predicting in
particular the vanishing of $M_{z}$ at either $B_{\rm c}$ or
$T_{\rm c}$. The incomplete saturation of $M_{z}$ at zero field
arises probably from 'down' spins that remain frozen below
$B_{\perp}=4$ T because the quantum tunneling rates become too
slow at such low fields. We face here the curious situation that
quantum fluctuations, which can eventually suppress magnetic
order, are nevertheless necessary to attain equilibrium. In Figure
\ref{phasediagram}, we show the magnetic phase diagram of
Mn$_{12}$ acetate obtained from our experiments. Application of a
perpendicular field tends to shift $T_{\rm c}$ significantly
towards lower temperatures. As before, the mean-field calculations
reproduce reasonably well the overall features.

Summing up, our experiments on Mn$_{12}$ nanomagnets show the
existence of a ferromagnetic phase below $T_{\rm c} \simeq 0.9$ K
that can be suppressed by the application of an external magnetic
field. Deciding if this field-induced transition is driven purely
by quantum fluctuations requires measuring the critical behavior
of $M_{z}$ \cite{Sondhi97}, which is clearly beyond the
sensitivity of the present experiments. What we do observe is
that, above $B_{\rm c} \sim 5.5$ T, the order parameter $M_{z}$
vanishes even at the lowest accessible temperatures ($T > 47$ mK
in our case). We note that $B_{\rm c}$ is about two thirds of the
anisotropy field $2SD/g\mu_{\rm B} \sim 9$ T that would be
required to saturate the Mn$_{12}$ spins along a hard axis if they
were {\em classical} spins. In fact, as shown in the inset of Fig.
2, $M_{\perp}$ is still much smaller than saturation at $B_{\rm
c}$. These facts and the agreement with mean-field calculations
make it plausible that quantum fluctuations can suppress
long-range order. Within this interpretation, $M_{z}$ vanishes
because the quantum tunnel splitting $\Delta$ induced by
off-diagonal terms of Eq. (\ref{Hamiltonian}) dominates over
dipolar interactions. Therefore, at $T=47$ mK almost all molecular
spins should be in their ground state, which becomes a
superposition of 'spin-up' and 'spin-down' states, a mesoscopic
magnetic "Schr\"{o}dinger's cat" \cite{Schrodinger35}. Previously,
the existence of quantum superpositions of spin states was derived
from the detection of $\Delta$
\cite{Luis00,DelBarco99,Bellessa99}, while here we have monitored
the vanishing of the $z$ spin component.

Similar collective magnetic phenomena could be observed in arrays
of larger nanomagnets, like magnetic nanoparticles, provided they
are sufficiently ordered and monodisperse, requirements that
appear to be within the reach of modern synthetic procedures
\cite{Sun00}. The existence of "super-ferromagnetism" was
predicted more than ten years ago \cite{Morup94}. However, besides
indications of the collective, spin-glass-like, nature of the
magnetic relaxation \cite{Jonsson95}, no clear-cut experimental
evidence for long-range order has been found yet. The present and
some other recent results \cite{Evangelisti04} show that bottom-up
synthesis can provide physical realizations of these
super-ferromagnets albeit on a smaller size scale. Furthermore,
quantum dynamics can be used to overcome the slow relaxation and
to switch between the ordered and paramagnetic phases.

The authors are grateful to J. L. Ragazzoni and D. Culebras for
technical assistance. We acknowledge useful comments on this work
by J. F. Fern\'andez. This work is part of the research project
MAT02-433 funded by Spanish MCyT. It was partly supported by the
European Commission under project IST-NANOMAGIQC. The authors
acknowledge financial support of the ILL for the preparation of
the samples.


\begin{thebibliography}{99}
\bibitem{neutron} Neutron diffraction
experiments were carried out on the diffractometer D10 at the
Institut Laue Langevin. We worked in a two-axis normal-beam
configuration at a neutron wavelength of $2.36$ \AA.
\bibitem{dipolar} Approximating Mn$_{12}$ by point-like spins gives
interaction energies within $2$ \% of the exact result only when
the intermolecular distances exceed $5$ lattice parameters. For
shorter distances, the disk-like shape of the Mn$_{12}$ molecule
reinforces dipolar interactions along the $a$ or $b$ axes with
respect to interactions along $c$. Following simple arguments
\cite{Fernandez00}, this should lead to an enhancement of $T_{\rm
c}$. The size and shape of the nanostructure becomes an additional
factor to determine (and eventually tailor) the physical behavior
of the array.
\end{thebibliography}
\end{document}